# A Storm in an IoT Cup: The Emergence of Cyber-Physical Social Machines

*Aastha Madaan[1], Jason R.C. Nurse[2], David De Roure[3], Kieron O'Hara[4], Wendy Hall[5] & Sadie Creese[6]*

**ABSTRACT:** The concept of 'social machines' is increasingly being used to characterise various socio-cognitive spaces on the Web. Social machines are human collectives using networked digital technology which initiate real-world processes and activities including human communication, interactions and knowledge creation. As such, they continuously emerge and fade on the Web. The relationship between humans and machines is made more complex by the adoption of Internet of Things (IoT) sensors and devices. The scale, automation, continuous sensing, and actuation capabilities of these devices add an extra dimension to the relationship between humans and machines making it difficult to understand their evolution at either the systemic or the conceptual level. This article describes these new socio-technical systems, which we term Cyber-Physical Social Machines, through different exemplars, and considers the associated challenges of security and privacy.

## 1. Introduction

While changing the way human communication and encounters take place (Berners-Lee 1999), the Web has itself changed from a transactional entity to a multitude of active spaces populated by agents both human and artificial autonomously interacting. These spaces bring various human actors in close proximity with each other, attracted by common ideas, interests, activities, and goals. This has led to formation of newer online communities and groups around various Web-based platforms, including social networking sites like Facebook (https://www.facebook.com/), messaging and microblogging sites like Twitter (https://twitter.com/), online resources of which Wikipedia (https://en.wikipedia.org/wiki/Main_Page) and Flickr (https://www.flickr.com/) are the most celebrated, and special purpose platforms such as citizen science platform Zooniverse (https://www.zooniverse.org/), music site last.fm (https://www.last.fm/), and healthcare sites such as PatientsLikeMe (https://www.patientslikeme.com/). These platforms are evolved ecosystems consisting of a variety of interactions and participation in co-creation by their end-users in a fluid range of capacities. Perhaps most dramatic is the case of Facebook, initially aimed at being a student directory but now used to run social and political campaigns, and hosting a whole zoo of groups, herds and mobs (Sivaraman and Srinivasa 2015). These human-platform interactions give rise to a number of ethical and governance challenges including privacy and security. Even though each of these platforms may be designed for some purpose, they may be used for others, with or without the blessing of their original designers, especially when people interact on the platform at scale (O'Hara et al 2013). This makes it both complex and interesting to study and understand them. However, the divergent nature of their evolution can only be studied by observing, analysing and replaying the analysis. Since the amount of data generated by the Web currently exceeds the capacity of scientists to process it, a sociotechnical platform called the 'Web Observatory' (Page and De Roure 2013, Tiropanis et al 2014, De Roure et al 2015, Tinati et al 2015, Madaan et al 2016) has been proposed to "observe, experiment and understand the Web" by making tools and methodologies available through it.

---

[1] Arup.
[2] School of Computing, University of Kent.
[3] Oxford e-Research Centre, University of Oxford.
[4] Electronics & Computer Science, University of Southampton, kmoh@soton.ac.uk.
[5] Electronics & Computer Science, University of Southampton.
[6] Dept. of Computer Science, University of Oxford.

In this paper, we take inspiration from Berners-Lee's (1999) definition of 'social machines' as abstract social engines on the Web that would allow people to create new forms of social processes and do creative work while machines do the administration, allowing new forms of social processes to emerge from technical affordances and constraints. This definition is normative, optimistic and flawed – not least because in many kinds of sociotechnical interaction, the humans do the routine stuff and the machine is creative – but it set the stage for a bigger revolution on the Web in which much human communication was displaced to platforms as described above, which appeared and disappeared according to their ability to host innovative and useful interaction (De Roure et al 2013). For example, Facebook and Twitter became indispensable and emerged as two of the largest platforms for human communication, while others such as MySpace and Orkut faded at their peak (Pachal 2018). Many platforms use other platforms as intermediaries; Ushahidi , a platform which organises citizen journalism via geospatial mapping, crowdsources information from activists using multiple channels, including SMS, email, Twitter and the Web (https://www.ushahidi.com/).

Social machines are inevitably becoming more social, more complex and more situated with the rapid adoption of Internet of Things (IoT) devices. IoT devices passively sense human activities, possess the capability to act on behalf of humans and are connected through (and on) the Web. While they bring a new kind of empowerment to people, they also require new ways to understand their impact on existing human-machine interaction, what Nissenbaum (2010) calls their contextual integrity. The automation, scale and actuation properties of these devices is inevitably leading to a paradigm shift in the affordances of social machines and giving rise to what we call Cyber-Physical Social Machines (CPSMs). In this paper, we describe a first generation of CPSMs, using exemplars to suggest how different types of CPSM might arise from a range of IoT affordances such as actuation, sensors, distributed networks, automation, and scalability. We reconceptualise the coupling between humans and machines in light of IoT devices and show how open issues facing social machines and the IoT itself (e.g., rigorous risk assessments and their adequate security – Roman et al 2013, Nurse et al 2017, Tinati et al 2017, Nurse et al 2018, N. Madaan et al 2018) are compounded in case of CPSMs. This is an intriguing line of research, considering that IoT not only contributes additional data points (related to the physical world, such as location) to social data, but also can automate certain interactions involving humans, platforms and devices.

The structure of the paper is as follows. Section 2 describes some exemplars of existing social machines. Section 3 then moves on to give some early and pioneering exemplars of CPSMs which follow from the wide-scale adoption of the IoT. Section 4 outlines specific concerns of privacy and security that revolve around CPSMs and section 5 concludes with a brief sketch of some open research questions.

## 2. Social machines

Certain kinds of interaction, based around platforms, sufficiently self-contained to understand as a coherent sociotechnical entity with a recognisable *telos*, and containing a measure of emergent social complexity (especially human-to-human communication) as well as technological sophistication, have been written about as social machines (Hendler & Berners-Lee 2010, Shadbolt et al forthcoming). Examples include social responses to modern problems of transport, such as Waze (https://www.waze.com/), a navigation app which uses community-derived real-time data about incidents such as traffic jams and accidents, and response to crime, such as BlueServo (http://www.blueservo.net/), which crowdsources policing on the Texas-Mexico border, and Onde Tem Tiroteio (Where the Shootouts Are – https://www.ondetemtiroteio.com.br/), which uses a network of a million people on Facebook, Twitter, Instagram, WhatsApp, Telegram and a special purpose app to provide real-time information about shootings and gang-related crime in Brazil. There is a growing number of social machines in the area of health (Van Kleek et al 2013) to pool patients' resources and to offer support and advice to fellow sufferers, such as PatientsLikeMe or curetogether (http://curetogether.com/). Social machines spring up to coordinate bottom up disaster response, such as the Cajun Navy (https://www.facebook.com/groups/TheCajunNavy/), in which leisure boat owners in Louisiana located and rescued victims of floods in 2016 via a Facebook group. Other social machines have developed for gaming, citizen science, the collection of data about, e.g. climate change or pollution, or exposing corruption.

Berners-Lee's broad concept has been extended and modified to encapsulate social machines' evolution, encompassing contingent developments in social networking, mobile devices and data production, among others. The following is a descriptive account that situates them firmly in the realm of social behaviour, and points to some of the dilemmas and opportunities that they create:

*A social machine can be defined metaphorically as a machine which has effects, actions, builds its own community and resources (data). These are then shared and consumed by the various stakeholders who themselves are the contributors. It is a process of human-system mutual co-production and involves social, digital, physical materiality. It exhibits emergent patterns of inclusion and exclusion. (De Roure. 2017)*

### 2.1   Exemplar #1: Social machines on the Web

Initially, the content and platforms on the Web were organised around the ways humans would read paper books, or papers. Web 2.0 paid attention to the relationship between the Web and human users through user-generated content, but did not specifically focus on direct human to human interactions (instead acting as an intermediary to facilitate human to human communication and create its well-known network effects). The next revolution led to a read-write social Web which supported the interactions among humans and those with the Web. The first set of social machines included highly expressive ones such as Wikipedia, and far less social entities, such as reCAPTCHA (Von Ahn et al 2008). In the case of Wikipedia, the interaction is mostly unidirectional where the human participants or bots edit the encyclopedia content and structure. In case of reCAPTCHA, the information may be fed back to participants to improve their experience or incentivize their participation, although its main purpose is to feed into the production of scientific knowledge (De Roure et al 2013).

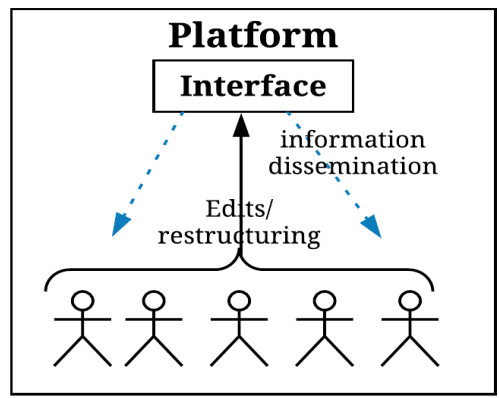

Figure 1: An illustrative example of social machines on the Web such as Wikipedia, and reCaptcha, which imitate human communication, perception and build communities coupled with the platform

Wikipedia is built on top of the MediaWiki (https://www.mediawiki.org/wiki/MediaWiki) software package which is used in several other applications that have not gained similar traction from such a broad community. Wikipedia's success is built on the way it supports article creation, edition and tracking in ways that mirror offline human problem-solving in the publishing world. In other words, the editorial process of Wikipedia is impacted by insights from the knowledge curated from human perception and communication, while being a paradigmatically Web-based system and humans can be viewed as 'embedded' within the platform with some of them also orchestrating processes for administering the site (Hendler et al 2008). In case of reCAPTCHA, not all its users understand its purpose (De Roure et al 2013). It is a machine aimed at harnessing human mental effort to preserve human knowledge available in archives to be able to digitise them. Most of the words presented in reCAPTCHA are images from books which cannot be deciphered using OCR programs (Von Ahn et al 2008). Therefore, reCAPTCHA is a social machine harnessing human mental effort without complex social interaction, and feeding it to a variety of social machines. Figure 1 illustrates this basic social machine ecosystem on the Web.

This simplified figure should not be taken to imply that social machines are standalone phenomena, to be treated methodologically as individuals. In one typical yet informative instance, described in (De Roure et al 2015), a social website which itself is a social machine was protected by another social machine (reCAPTCHA), but despite this came under spam attack. The attack linked the spammers with the social machine's ecosystem, but interestingly from this perspective the spammers were themselves coordinated by new social machines farming the spam; hence the new addition to the social machine's ecosystem was itself a social machine (or perhaps an anti-social machine). To combat the problem the original social machine's administrators created yet another new social machine, transforming the original machine to make use of another social machine for flagging spam and blacklisting spammers. This example illustrates aspects of multiple interacting social machines reacting with or against each other, the ecosystem informing their design and composition.

## 2.2   Exemplar #2: Social machines on handheld devices

Twitter, Facebook and Zooniverse are important examples of platforms designed for a given purpose which evolved to accommodate imperatives in human communication and behaviour through software features such as 'likes' on Facebook or 'hashtags' on Twitter. These features act as handles for expressing and propagating ideas and opinions across different communities, and enable straightforward quantification which provides extra affordances that can be furnished by software. As shown in Figure 2, human communication and opinion sharing on these platforms, with these novel features, create strongly coupled communities and groups of people sharing goals, tasks, ideas and opinions.

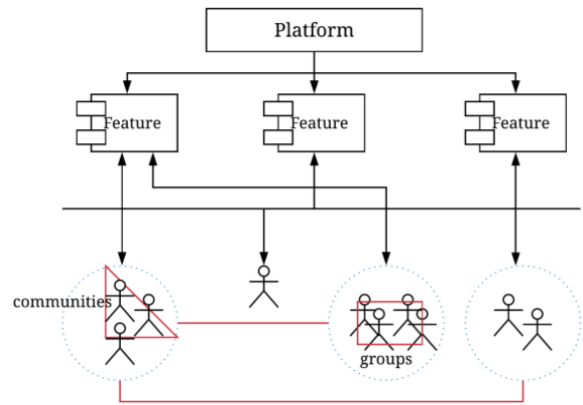

Figure 2: An illustrative example of social machines which sense human behaviour through human participation, communication and other social characteristics exhibited by them through use of available platforms whose features actively influence their opinions and decisions. Humans are tightly coupled not just to the system but also to each other in these social machines.

With the widespread adoption of mobile devices, these platforms came up with lightweight applications that could create these couplings away from the PC. GPS may seem to be merely an additional data point, but has revolutionised these applications, allowing greater personalisation for users based on sensing their location from their devices. In turn it provides new means of abusing a social machine through techniques such as GPS spoofing. The constant connectivity of mobile phones allows people to interact and collaborate in their preferred settings in real time (Roush. 2005). Mobile devices enable more participants to engage more continuously, and this scaling works even for platforms that involve expert intervention if the participants are diverging from their nominal task. Examples include MOOCs, and the citizen science platform Zooniverse; these deliberately influence participants' initial strategies for problem-solving and support expert intervention if the participants are diverging from their nominal task, yet even they engage more participants more continuously through their mobile versions (Tinati et al 2017).

## 3. Cyber-physical social machines

Advances in computing, storage, Internet access, wireless communication, and sensing have made it possible to monitor and analyse human behaviour, social interactions, and city dynamics on a large scale and in near real time (Gyrard et al 2016). Until the development of digital communications technology, social and demographic data had to be deliberately collected by governments, corporations and academics, often through surveys designed to illuminate a particular issue, but also via analysis of data created as a by-product of very specific interactions (for instance, applications for government welfare, or complaints about a product). Data was usually created about social phenomena, and only in specialised circumstances through social behaviour. Social data now is almost all created through behaviour, driven not by the needs and functions of governments and corporations, but by the voluntary actions of individuals and groups. The communications channels themselves are designed to be increasingly valuable to individuals as the size of their networks grows, while simultaneously valuable to their creators by automatically making a record of the key aspects of the communication. This has the effect of (a) producing a much finer-grained portrait of human interaction, that (b) is not directly connected to the business concerns of the platform (and therefore much more likely to be surprising and interesting), and that (c) can be fed back to the interacting individuals, or (d) can be used to adjust the context of the interaction. Social machines succeed because of the rich (and manipulable) picture of communication that social data provides. Currently we still tend to still work with 'found data' or digital exhaust, rather than influencing the data collection. However, technology is increasingly giving us the affordances to engage in the design of data collection rather than (merely) recycling emissions.

The IoT is modifying our physical world and our interaction with cyberspaces, from ways we remotely control appliances at home to the way care is provided to the elderly (Gyrard et al 2016, Smart et al 2017). It is extending this fine-grained record beyond communication to include physical activity and movement within the physical environment, thereby removing one more level of abstraction from the data representation of individuals, representing them not only as social individuals, but also as individuals in a physical context. Smart et al (2017) described expansion of social machines research to encompass the IoT with the viewpoint of being able to isolate a number of points of contact between the science of social machines and the sciences of the embodied mind. The following characteristics of the IoT are key in this context:

- **Scale.** The connectivity of IoT devices to the Web and the Internet allows them to scale as a distributed network of networks. This can create a strong local coupling among systems and loose coupling with external systems (Nurse et al 2017, N. Madaan et al 2018).
- **Automation.** Autonomous data collection and processing, making contextual inferences (Sarkar et al 2014), collaborating with other IoT platforms and services, and taking action based on real-time inputs are key to newer social machines that will flourish within IoT-enabled contexts such as smart homes and smart cities.
- **Sensory capabilities.** Sensors are the core of IoT technology which collect data and information (Little 2017). They are capable of sensing peoples' environment to develop a context around them.
- **Actuation.** Actuators are an important component of an IoT ecosystem such as a smart home, or industrial IoT. From a social machine perspective, actuators complete the feedback loop and take decisions according to user-defined specifications.

We are now in the realm of cyber-physical social machines (CPSMs), which spring from the flourishing of communities around devices (such as Fitbits, https://www.fitbit.com/, Amazon Echo, https://www.amazon.co.uk/gp/product/B06Y5ZW72J, and others) which interact with platforms. When platforms like IFTTT (If This Then That – https://ifttt.com/discover), Microsoft Flow (https://flow.microsoft.com/en-us/) and Zapier (https://zapier.com/apps/integrations) are introduced to enable people to integrate their personal and environmental devices with their social media and other Web services, they become inclusive of communities around the devices they support, creating more heterogeneous ecosystems with more diverse and broader groups of participants. These communities or groups can control and innovate communication to their devices in ways not always anticipated *a priori*. Hence these platforms can be considered as ecosystems which are social in a fundamental sense.

## 3.1   Exemplar #3: Social machines as sensors

Smart trackers and wearables such as Fitbit actively sense our physical activity and sleep routines. In addition they provide a free mobile application that syncs the data sensed by the device, and allows people to share their data on social media platforms and also with other friends using similar devices. Within the same application one can observe one's personal physical activities on a daily basis but also connect to communities around wellbeing. One can also connect to social media platforms and share fitness goals and achievements (through badges in the case of Fitbit). This has eliminated the need for people to carry their mobile devices during their run and often log their run offline with different applications such as Fitbit's mobile app (https://goo.gl/esvFgi) or Runkeeper (https://runkeeper.com/).

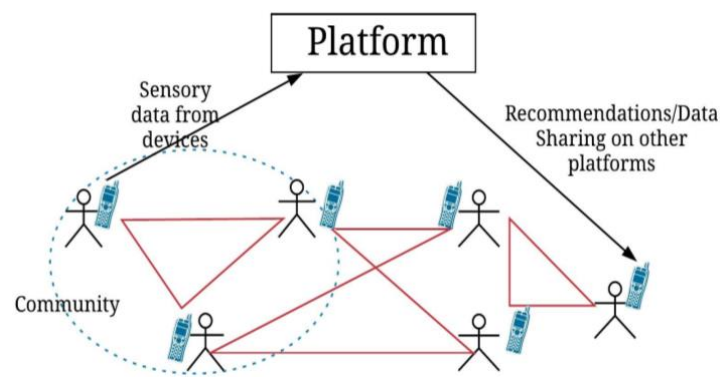

Figure 3: An example illustrating how sensory data is captured by a mobile based IoT platform. These users form active communities of participants with similar interests, who share data, and take part in joint activities.

This fitness CPSM senses people's personal fitness parameters and connects them with other fitness-concerned friends and communities in a variety of ways. As shown in Figure 3, the different users having a sensory device such as a mobile phone or fitness tracker contribute their data to a platform, and in return the platform gives advice or recommendations, e.g. of health plans, and connects them to other people with similar interests or who own a similar device. Next, these people start interacting through notifications and messages familiar from online social media platforms. It is important to note here that the devices share data and statistics in an automated manner with the network a particular user has chosen. In addition, this platform connects participants to other platforms and their users, bridging physical remoteness. Therefore, we observe that the communities around devices merge creating more complex ecosystems between platforms as well as between the participants.

## 3.2   Exemplar #4: Social machines as actuators

Web-based platforms such as IFTTT help users connect their applets and services together. These include physical devices as well as digital assistants such as Cortana (https://www.microsoft.com/en-gb/windows/cortana) or Amazon Alexa (https://en.wikipedia.org/wiki/Amazon_Alexa). For example, one can set up a 'connected home' with the digital assistant; users need only open Cortana on Windows 10 or go to the Cortana application on their phone, click Notebook and then click 'Connected Home' (Martin and Finnegan 2018). Users can create their own applets by combining various applet 'services' and setting trigger parameters. In addition applets created by a user can be shared and used by another user on her device. This creation and consumption between the communities around these platforms resonates with the social machines characterisation given in Section 2.

An increasingly popular way to use IFTTT is in conjunction with the Alexa voice assistant. A number of use-cases centre around familiar IoT applications such as controlling smart home devices with voice commands directed at Echo and Echo Dot speakers (telling Alexa to make a pot of coffee with a WeMo-connected coffee maker, or changing the colour of Philips Hue smart lights each time Alexa plays a new song – Martin and Finnegan 2018). Therefore, once the user configures the trigger parameters and sets up the services, she is taken out of (or retires from) the loop and the devices talk to each other and the underlying platform. This new kind of ecosystem can be considered as a social machine in which humans are not concerned with the quotidian detail, while a community of devices interacts autonomously, sharing data and performing actions. Human sociality becomes far less complex, even as the interactions become more tightly coupled.

Figure 4 describes the tight coupling that occurs between devices in these CPSMs. Though the connected sensing and actuating devices belong to different platforms, their interconnection provides the ability to share information across platforms through a unified framework, developing a common operating picture for enabling innovative applications across a network of users, applications and devices (Gubbi et al 2013).

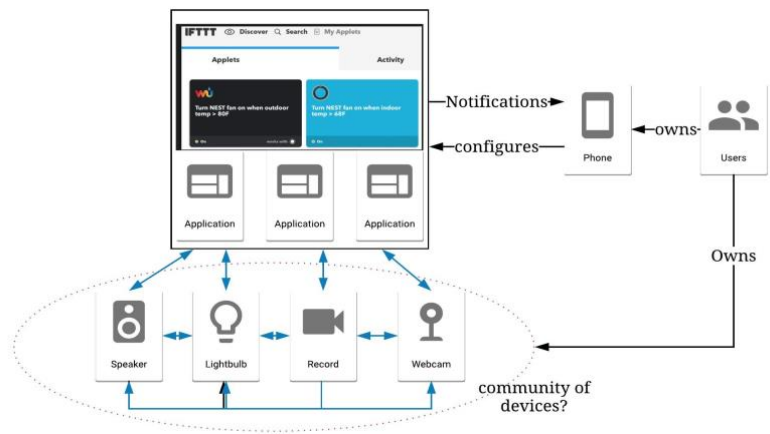

Figure 4: An illustrative example of a CPSM, where humans are taken 'out-of-loop', and devices are connected to a platform which supports platform-to-platform and device-to- device communication, actuation and sensing.

## 4. Observing cyber-physical social machines

Smart homes and cities currently provide more of a reCAPTCHA-style world, focusing on automation, but we expect humans to build new social machines over this infrastructure – like Wikipedia over Mediawiki (Section 2.1) with variable success. The debate will then be about the empowerment afforded by the smart home or smart city to build CPSMs using platforms such as IFTTT, creating automated social machine ecosystems, where 'traditional' Web-based social machines will meld with physical infrastructure such as personal assistants. This can be described by a coupling and network effect, in which loose offline coupling of humans is tightened with close proximity on social media and the Web, and narrowed down further by IoT. While social machines had social complexity with the humans in the loop, CPSMs bring in automation which retains a (simplified) sociality, while the humans are mostly taken out of the loop.

We make no claims in this paper that this development will be necessarily good or bad; we merely mark its increasing presence. It will no doubt afford the potential for the empowerment of individuals and communities, but may also increase the potential for corporations to appropriate unpaid human labour (Ekbia & Nardi 2017), or for governments to direct (or 'nudge') human behaviour or to place it under surveillance. Time, and good design, will tell.

CPSMs will bring inevitable challenges of governance and ethics since they involve humans and their resources. The IoT brings with it threats of privacy-violating interactions, life-cycle transitions, inventory attacks and information linkage (N. Madaan et al 2018) arising during data aggregation and standardisation phases (Ziegeldorf et al 2014, Nurse et al 2017, Williams et al 2017). Another hard to predict factor is that of emergent behaviours in complex systems (O'Hara et al 2013), including the accidental assembly of CPSMs; the opportunity to innovate always needs to be balanced against unintended consequences, especially in these hyper-social environments at scale.

## 4.1  Security

As technology in general has advanced, there has always been a need to consider security and its foundational tenets of confidentiality, integrity and availability of data and services. CPSMs contain two central components that are common targets for a range of cyber-attacks – human participants and platforms. Furthermore, the IoT devices themselves have vulnerabilities and security challenges which open them more to attack (O'Hara 2014, Nurse et al 2015). Therefore, CPSMs are especially vulnerable to the increasing wave of attacks that aim to exploit platform technologies and specifically target users. Social engineering and phishing attacks can severely impact any of the exemplar social machines. This could be either whilst the victim is participating in the machine or is an initial developer of the platform. While these threats apply broadly, Exemplars #3 (section 3.1) and #4 (section 3.2) are more vulnerable because control lies with the devices.

In addition to considering the increased attack surface presented by CPSMs, there are other security susceptibilities that need to be addressed. For example, with respect to Exemplar #4, as smart homes, offices and cities become more connected, the participating actuators, hackers or even hacktivists could use them to launch cyberattacks meant to disrupt or directly harm individuals, neighbourhoods and enterprises. This goes beyond hijacking voice assistant technologies, and could mean hackers enabling machines remotely, compromising the confidentiality or integrity of services, or conducting denial-of-service attacks against actuators. The connected and dynamic nature of the IoT also can lead to a propagation of such harms instantaneously across systems (Nurse et al 2017, Nurse et al 2018). This is exacerbated by the sociality of social machines, relationships and interactions across these machines. Another critical concern is security risk assessment which is largely an unsolved issue in the IoT (Nurse et al 2018). In case of CPSMs, different social machines have a variable coupling between human participants and the underlying platforms. To combine the two areas, risk assessment would need to take into account the objects of the CPSMs (e.g. the intended output, and degree of engagement) and consider the consequence to these objects in the event of a malign interference of some sort (an intrusion or an insider). This may be particularly challenging when CPSMs grow in ways unknown *a priori*.

## 4.2  Privacy

Privacy is another significant concern, which needs to be addressed through an interdisciplinary approach. The increasingly fine-grained picture of individuals provided by social data and IoT data clearly threatens privacy to a greater extent than known previously (Sicari et al 2015). In the first place, it contains more information about individuals at a less abstract level, and consequently there is greater scope for a breach of privacy. Multiple examples can be found in the media today of breaches in new forms of technology including the IoT having a severe privacy impacts on individual users (O'Hara 2014). For our vantage point, this is especially important given the sensory capabilities of CPSMs.

Secondly, communication using a digital channel is much more likely to follow an informed choice on the part of the communicator (and hence can be consented to meaningfully) than moving through an instrumented physical environment, especially when the environment is a public space. It may be that people are relatively relaxed about the privacy implications of the IoT, but whereas it is straightforward (if controversial) to argue that someone who uses a social machine such as Facebook has consented to the use of her data in various ways (even if the consent is not very informed, the person does have the choice of exit, however painful, in the event of concern), it is far harder to make a similar case about someone who merely inhabits a space. If the space is private property, such as a shopping mall, then it might be argued that consent for the use of the data follows implicitly from the fact of entry; even this is an open question. If the space is public commons however, then consent to the gathering and processing of IoT data is surely absent in the general case.

As we focus on personal data and privacy implications in CPSMs, numerous other important questions arise. For instance, how do individuals understand and perceive privacy in situations where CPSMs regularly, seamlessly and automatically collect, aggregate and share personal data? Even in cases where participants are privacy-aware and able to express their preferences, how can data be shared across social machines such that it preserves each participant's individual requirements and desire for privacy? While better understood and more mature social machines (for instance, those in Exemplars #1 and #2 above) can be designed with answers to these questions in mind, the understanding of privacy and respective privacy capabilities are not widely present in IoT environments. In many ways, privacy in these cases is still unfortunately more of an afterthought (Williams et al 2017) as are the risks related to data aggregation (Aktypi et al 2017) and meta-data leakage (Nurse et al 2014, N. Madaan et al 2018).

## 5. Conclusion: research challenges

Reflecting on these exemplars of social machines and CPSMs, we can see that many meaty research challenges face the IoT community. From a methodological point of view, both the intersection and parallels of social machines and IoT (inclusive of their communities and platforms) need to be considered. Similar to the approach of observing social machines and their ecosystems through a social machine observatory (De Roure et al 2013), we can take an observatory approach to understand the social aspects of various existing and emerging CPSMs. This would suggest an IoT Observatory (A. Madaan et al 2018) which observes data, resources and human-machine interactions in various IoT ecosystems, provides technical features and grounds the social aspects in theory. Other concerns that need to be researched include:

1. *Tight coupling and close proximity* within the distributed network of heterogeneous actors and platforms need to be analysed using qualitative and quantitative studies to understand

   the intended purpose and actual use. For social machines, Tarte et al (2014) described the role of stories, and used archetypal templates from prosopography (2015) to create narratives for describing the coupling between the underlying technological platform and the social layer (of participants).
2. *Provenance summaries* can play a critical role in CPSMs, especially in models where the degree of automation is high. This could support discovery of emerging communities, quantify the degree of overlap between ecosystems of social machine(s) with CPSMs and understand the degree of control and participation of its users.
3. Following De Roure et al (2013) we need to establish the *ground rules* that lead to the emergent behaviour: these may be explicitly stated as rules by which people abide, could be encoded into the technology, or may potentially be norms of community conduct or grounded in how other CPSMs influence behaviour.
4. *Security and privacy threats* are a critical aspect of CPSMs due to their continuous online connectivity and availability. In particular, there is an increased likelihood of attack given the openness of such machines, and a larger impact when we consider their interconnected nature, dynamism and changing elements of risk. Therefore, research is required to distinguish and understand the security and privacy threats that are created for (and by) social actors and platforms within these ecosystems. Systems are increasingly coupled, as social media meets cyber-physical.

This paper has drawn on the theory and practice of social machines to understand the emerging cyber-physical social machines, an important area of research which has fallen between the cracks of social IoT and social computation. Where social machines and IoT intersect, the social data from social machines will be supplemented with additional data points about people's physical activity and context, extending the social machines to a physical dimension. Our study sketches likely structures of first generation CPSMs based on the sensory and actuation capabilities of IoT devices. It describes the relationship between the IoT, Web and human participants, the key characteristics of degree of interactions, coupling and automation and their socio-technical implications. We must further emphasise the challenges of security and privacy that may arise due to both the social embeddedness and accessibility of devices over the Web, as well as many other open research questions.